\documentclass[10pt,
twocolumn,
amsmath, 
amssymb, 
groupedaddress,?/
aps,
prl,
nofootinbib,
raggedbottom,
longbibliography]{revtex4-2}

\usepackage{graphicx} 
\usepackage{color}
\usepackage[dvipsnames]{xcolor}
\usepackage{cancel}
\usepackage{tikz-cd}
\usepackage[justification=raggedright,singlelinecheck=false]{caption}
\usepackage{subfiles}
\usepackage[percent]{overpic}

\usepackage{amsmath, amsfonts, amssymb}
\usepackage{mathtools}
\usepackage{bbm} 

\usepackage[markup=underlined]{changes}
\definechangesauthor[name=Tin, color=blue]{ts}

\usepackage{import}

\usepackage{transparent}

\usepackage{caption}
\DeclareCaptionJustification{fulljustify}{\leftskip=0pt \rightskip=0pt \parfillskip=0pt plus 1fil}
\renewcommand{\emph}[1]{\textit{#1}}

    \usepackage{tikz}

\newcommand{\diff}{\mathrm{d}}

\newcommand{\avg}[1]{\left\langle#1\right\rangle}

\begin{document}
\title{Proof of entropic order in Generalized Ising Models \\}
\author{Enrico Andriolo}
\author{Mendel Nguyen}
\author{Emily Richards}
\author{Tin Sulejmanpasic}
\affiliation{Department of Mathematical Sciences, Durham University, Durham DH1 3LP, UK \looseness=-1}

\begin{abstract}
Ordering at arbitrarily high temperature---entropic order---has been argued to take place in a class of generalized Ising models parameterised by a real interaction parameter $p$ when $p\ge 1$. We give a rigorous proof of this conjecture. We further show that on arbitrary graphs, these models solve graph packing problems---crucially, the Maximum Independent Set optimisation problem. Due to the NP-hardness of this packing problem on generic graphs, some lattice systems will exhibit glassy phases. We call this phenomenon \emph{entropic glass}.

\end{abstract}

\maketitle
\newpage

\section{Introduction}

As systems are heated, thermal states favor high entropy and therefore typically disorder. While ordered phases can exist at intermediate temperatures as in the Pomeranchuk effect \cite{pomeranchuk1950theory,Richardson1997}---a phenomenon called inverse melting---they are not robust against infinite heating when the temperature exceeds the typical microscopic interaction scale.\footnote{Models with infinite repulsion \cite{Alder1957,Onsager1949,Frenkel2015,Schupper_2005,baxter1980hardsquare} are of course resilient to heating, because the interaction scale is infinite.} Indeed, there are rigorous theorems which state that disorder must be restored at sufficiently high temperature \cite{dobrushin1968description,kuntsch1982decay,friedli2017statistical, Weinberg1974}.

Recently, it has become clear that the assumptions of such theorems may be too stringent; there may exist physically plausible systems which violate them and order up to arbitrarily high temperature \cite{Chai:2020onq,Chai:2021tpt, Chai:2021djc,Hawashin:2024dpp,Komargodski:2024zmt,Han:2025eiw,Hawashin:2025ikp,Huang:2025gvi}.

In particular, a class of generalized Ising models given by the Hamiltonian
\begin{equation}
    H[n] = U \sum_{\avg{ij}}n_i^p n_j^p+\mu\sum_i n_i,
    \label{hamiltonian}
\end{equation}
was argued to exhibit ordering at high temperature. This phenomenon is called entropic order \cite{Han:2025eiw}. In the above Hamiltonian, $U>0$ and $\mu>0$ are parameters with dimensions of energy, while $p>0$ is dimensionless. The ``spins'' $n_i=0,1,\cdots$ can take arbitrarily large integer values. 

Mean field theory (MFT) analysis and Monte Carlo simulations indicate that this classical system orders for arbitrarily high temperature \cite{Han:2025eiw} when $p>1$. 
Ordering occurs because entropy prefers large $n_i$ fluctuations. 

Indeed, MFT demands that in a gas phase $n_i\sim T^{1\over 2p}$ for $T\gg U,\mu$. On the other hand, on a bipartite lattice (e.g. square), one of the sub-lattices ($A$) may have spin values of $n_i\sim T$ as long as the other sub-lattice ($B$) is entirely depleted, thereby increasing the total entropy (as long as $p>1$) at the expense of spontaneously breaking the lattice translational symmetry for arbitrarily high $T$.

Furthermore, for the marginal case $p=1$ it was shown \cite{Huang:2025gvi} that high-$T$ order is favored as long as $U>U_c$ for some fixed $U_c$. While the value of $U_c$ was not established exactly, a ``large color'' model indicates that the mean-field value of $U_c=\mu/2$ may be exact \cite{Huang:2025gvi}. 

Despite insights from MFT and Monte Carlo simulations, neither program can rigorously establish the existence of order for arbitrarily high temperature. 
In particular, MFT is plagued by large temperature fluctuations beyond theoretical control.\footnote{Despite this, MFT and Monte Carlo simulations agree quite well, which suggests that the offending MFT fluctuations may cancel order by order \cite{Huang:2025gvi}.}
Here, we prove that order exists at arbitrarily high temperature for $p>1$ on any bipartite lattice in two or more dimensions. To our knowledge, this is the first proof of entropic order in a lattice system.

\begin{figure}[b]
    \centering
    \def\svgwidth{.8\columnwidth}
    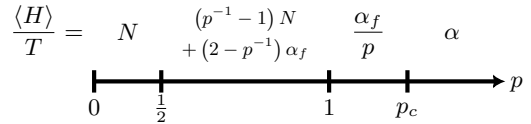
    \caption{A sketch of different regimes of the Hamiltonian \eqref{hamiltonian} on an arbitrary graphs for $p\geq 0$. $\alpha$ and $\alpha_f$ are sizes of the MIS and MFIS respectively (see text). For general graphs, a single transition at $p_c$ may be replaced by multiple, intermediate regimes solving a complicated optimization problem. Only when $p<1/2$ does the high-$T$ regime behave as $N=|G|$ independent degrees of freedom, as would na\"ively be expected for a high-$T$ disordered gas phase.}
    \label{fig:equipartition}
\end{figure}

We then generalize our methods to any arbitrary graph and discover that for $p$ sufficiently large, the dominant high-$T$ phase corresponds to populating vertices which belong to a \emph{maximum independent set} (MIS) of the graph. We call this the MIS-solid phase. On the other hand, for $p\sim 1$ the phase is governed by the \emph{maximum fractional independent set} (MFIS) of the graph. We call this the MFIS-gas phase.\footnote{Despite the name, the MFIS phase may not always be a gas phase. In contrast, the MIS phase is always a solid.} The MIS and MFIS are packing problems in graph theory which, as we will review below, are characterized by optimal sizes $\alpha$ and $\alpha_f$ respectively. These sizes are reflected in the high temperature equipartition theorems for each phase, summarized in Fig.~\ref{fig:equipartition}.

Finally, while the MFIS optimization problem can be computed in polynomial time, the MIS optimization is known to be NP-hard \cite{Karp1972} on a generic graph. This suggests that the corresponding phase is glassy. We dub such phases \emph{entropic glass}.

\section{Activation Variables and Their Weights}

The lattice is taken to be a generic graph $G$. We will find it useful to explicitly indicate the $G$-dependence of the Hamiltonian \eqref{hamiltonian} by writing $H_G$. 

\begin{figure}
    \centering
    \def\svgwidth{.85\columnwidth}
\begingroup%
  \makeatletter%
  \providecommand\color[2][]{%
    \errmessage{(Inkscape) Color is used for the text in Inkscape, but the package 'color.sty' is not loaded}%
    \renewcommand\color[2][]{}%
  }%
  \providecommand\transparent[1]{%
    \errmessage{(Inkscape) Transparency is used (non-zero) for the text in Inkscape, but the package 'transparent.sty' is not loaded}%
    \renewcommand\transparent[1]{}%
  }%
  \providecommand\rotatebox[2]{#2}%
  \newcommand*\fsize{\dimexpr\f@size pt\relax}%
  \newcommand*\lineheight[1]{\fontsize{\fsize}{#1\fsize}\selectfont}%
  \ifx\svgwidth\undefined%
    \setlength{\unitlength}{328.67198794bp}%
    \ifx\svgscale\undefined%
      \relax%
    \else%
      \setlength{\unitlength}{\unitlength * \real{\svgscale}}%
    \fi%
  \else%
    \setlength{\unitlength}{\svgwidth}%
  \fi%
  \global\let\svgwidth\undefined%
  \global\let\svgscale\undefined%
  \makeatother%
  \begin{picture}(1,0.41505613)%
    \lineheight{1}%
    \setlength\tabcolsep{0pt}%
    \put(0,0){\includegraphics[width=\unitlength,page=1]{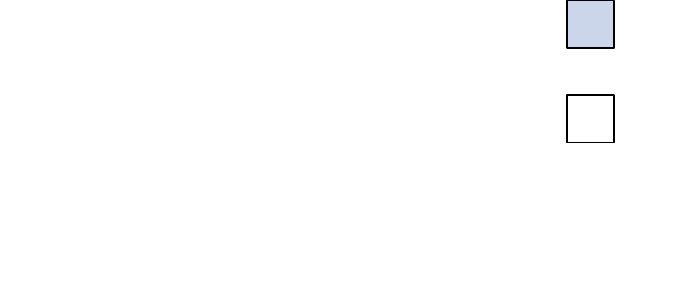}}%
    \put(0.90460091,0.22967991){\color[rgb]{0,0,0}\makebox(0,0)[lt]{\lineheight{1.25}\smash{\begin{tabular}[t]{l}$s_i=0$\end{tabular}}}}%
    \put(0.90460091,0.3676726){\color[rgb]{0,0,0}\makebox(0,0)[lt]{\lineheight{1.25}\smash{\begin{tabular}[t]{l}$s_i=1$\end{tabular}}}}%
    \put(0,0){\includegraphics[width=\unitlength,page=2]{induced_graph.pdf}}%
  \end{picture}%
\endgroup%

    \caption{An illustration of how the activation variables induce a subgraph $\widetilde G$ on the square lattice $G$. 
    }
    \label{fig:induced_subgraph}
\end{figure}

The analysis of the system is simplified by the introduction of activation variables $s_i$, which take values $s_i=0$ when $n_i=0$ and $s_i=1$ otherwise.
These variables pick out the set of vertices on which the occupation number $n_i$ is nonzero and induce a subgraph $\widetilde G \subset G$ which consists of the occupied vertices and all edges connecting them (see Fig.~\ref{fig:induced_subgraph}).

This allows the partition function to be re-expressed as a sum over all such vertex-induced subgraphs $\widetilde G$:
\begin{equation}\label{eq:partition_function_into_sectors}
    Z =\sum_{\widetilde G \subset G} W[\widetilde G],\qquad W[\widetilde G]=\left(\prod_{i\in \widetilde{G}} \sum_{n_i=1}^\infty\right) e^{-\beta H_{\widetilde{G}}[n]},
\end{equation}
where $\beta=1/T$ is the inverse temperature.
The weight $W[\widetilde G]$ factorizes into the product of the weights of the connected components $C$ of $\widetilde G$, $W[\widetilde G]=\prod_{C}W[C]$. We can therefore focus our attention on understanding the temperature-scaling of the weight for a single connected component, $W[C]$.

If $C$ is a simple isolated vertex, i.e. $|C|=1$, then:
\begin{equation}
    W[C]=\sum_{n_i=0}^\infty e^{-\beta\mu n_i}=\frac{1}{e^{\mu\beta}-1} \stackrel{T\gg\mu}{\approx} \frac{T}{\mu}\quad.
    \label{isolated_vertex_asymptotic}
\end{equation}
When $|C|\ge 2$, finding the asymptotic scaling of $W[C]$ as $T\rightarrow\infty$ becomes more complicated. To proceed, we first approximate the sum \eqref{eq:partition_function_into_sectors} by an integral $\widetilde{W}[C]$

\begin{equation}
    W[C]\approx\widetilde W[C]\equiv \left(\prod_{i\in C} \int_1^\infty d n_i\right) \;e^{-\beta H_{C}[n]}.
    \label{integral_weight}
\end{equation}
This is naively justified by the replacement $y_i=\beta n_i$, which allows one to rewrite the sum as a Riemann integral $\sum_{{n_i}}=\frac{1}{\beta}\int \diff^N y$ in the limit of small $\beta$. In the Supplemental Material \cite{SM}, we show that the integral $\widetilde{W}[C]$ has indeed the same asymptotic behavior as the sum $W[C]$ in the $\beta\rightarrow 0$ limit, up to a graph-dependent, but $\beta$-independent, prefactor. This will be sufficient for our purposes.

To evaluate \eqref{integral_weight}, we perform the following change of variables:
\begin{equation}
    n_i=\left(\frac{T}{U}\right)^{x_i/p},\qquad x_i\ge0.
\end{equation}
This results in the expression
\begin{multline}\label{eq:WtildeP}
\widetilde W[C] = \frac{\lambda^N}{p^N}\biggl(\prod_{i \in C} \int_0^\infty d x_i\biggr)\; \exp\biggl({\frac{\lambda}{p}\sum_i x_i}\biggr) \\ 
\times \exp\biggl({-\sum_{\avg{ij}}e^{\lambda(x_i+x_j-1)}-\frac{\mu}{U}\sum_i e^{(x_i/p-1)\lambda}}\biggr),
\end{multline}
where $\lambda=\ln (T/U)$. Crudely speaking, when $T\rightarrow\infty$ (i.e. $\lambda\rightarrow\infty$) the integrand is super-exponentially suppressed except in the region
\begin{equation}\label{eq:Pdomain}
    \mathcal A(C) = \{ \vec x \in [0,1]^N \;:\; x_i+x_j\le 1 \;\; \forall \avg{ij}\in C\}.
\end{equation}
Once the integral is restricted to this region, the integrand simplifies to $\exp({\frac{\lambda}{p}\sum_i x_i})$. We therefore expect the dominant contribution to the integral in the limit $\lambda\to\infty$ to come from the maximal value of the linear function $\sum_i x_i$ subject to the constraints \eqref{eq:Pdomain}.

The maximum value of the linear function $\sum_i x_i$ for $x_i$ within \eqref{eq:Pdomain} corresponds to a well-studied notion in graph-theory and linear programming. It is the size of a \emph{Maximum Fractional Independent Set} (MFIS) of $C$ and is usually denoted as $\alpha_f(C)$.

Let us first discuss the \emph{Maximum Independent Set} (MIS). An independent set of a graph $G$ is any subset of its vertices such that no two vertices are connected by an edge of $G$. The \emph{maximum} independent set $G_{\rm MIS}$ (not necessarily unique) is the largest of all such subsets, and the size of any such $G_{\rm MIS}$ is denoted $\alpha(G)=|G_{\rm MIS}|$. 
The problem of computing $\alpha(G)$ can be formulated as the assignment of an integer $x_i=0,1$ to every vertex $i$ such that $\sum_i x_i$ is maximized subject to constraints $x_i+x_j\le 1$ for all $\avg{ij}\in G$. For generic graphs, this problem is NP-hard \cite{Karp1972}.

A related optimization problem relaxes $x_i$ to take any real values $x_i\in[0,1]$ and admits a polynomial-time algorithm. The optimization problem reduces to a straightforward linear programming problem and the optimal value $\alpha_f(G)=\max \sum_{i\in G}x_i$ is called the size of the \emph{maximum fractional independent set} (MFIS) of $G$. As before, solutions $\{x_i\}$ need not be unique, and in fact can form a continuous moduli space of solutions that we denote $\mathcal{M}_f(G)$. By definition, the space of MFIS solutions must contain the discrete MIS solutions, and therefore $\alpha_f(G)\ge \alpha(G)$. The ratio $p_c=\frac{\alpha_f(G)}{\alpha(G)}$ is called the integrality gap. When $p_c=1$, $\alpha(G) = \alpha_f(G)$ can be computed in polynomial time.

Now it becomes clear that 
\begin{equation}
    W[C]\sim\widetilde W[C]\sim e^{\frac{\lambda}{p}\max(\sum_i x_i)}\propto T^{\alpha_f(C)/p},
\end{equation}
when $T$ is large. This is one of our main results, which will allow us to analyze the systems at high $T$. We give the main ideas of this proof in the following section, leaving the rigorous treatment to \cite{SM}. 

\section{Asymptotic Formula---main ideas}
Suppose $\vec X$ is a solution to the MFIS problem associated with $C$; i.e. $\sum_{i=1}^NX_i=\alpha_f(C)$, where $N=|C|$. 
Changing integration variables to $\vec y$ defined by $\vec x = \vec X + \vec y/\lambda$ gives
\begin{multline}
\label{integral_y}
    \widetilde W[C]=\frac{1}{p^N}\biggl(\frac{T}{U}\biggr)^{\alpha_f(C)/p} \biggl[\prod_i\int_{-\lambda X_i}^\infty d y_i\biggr] \\ 
    \times \exp \biggl( \frac{1}{p}\sum_i y_i - \sum_{\avg{ij}} e^{- \lambda s_{ij}}e^{y_i+y_j} \\
    -\frac{\mu}{U} \sum_i e^{-\lambda(1-X_i/p)}e^{y_i/p} \biggr),
\end{multline}
where $s_{ij}=1-X_i-X_{j}\ge0$. If $s_{ij}=1$ on the entire moduli space of solutions $X_i$, we call the bond $\avg{ij}$ \emph{active}.

The rigorous treatment proceeds to bound $\widetilde W[C]$ from above by discarding the $\mu$ term, as well as contribution from bonds which are not active. Similarly, one can bound $\widetilde W[C]$ from below by restricting the integration over $x_i$ to the region $\mathcal A(C)$ defined in \eqref{eq:Pdomain}. In this latter case, we already saw that that the integral behaves as $\sim T^{\alpha_f(C)/p}$, so we shall focus our attention on the upper bound.

Therefore, we have that
\begin{multline}\label{eq:tildeW_naive}
    \widetilde W[C]\le \frac{1}{p^N}\left(\frac{T}{U}\right)^{\alpha_f(C)/p} \biggl[ \prod_i \int_{-\infty\cdot X_{i}}^\infty dy_i \biggr]\\ \times \exp \biggl( \frac{1}{p} \sum_i y_i -\sum_{\avg{ij} \, \rm{active}}e^{y_i+y_j} \biggr),
\end{multline}
where we also took the limit $\lambda\rightarrow\infty$ in the integration limits (the notation $\infty\cdot X_i$ means $\infty$ if $X_i \neq 0$ and $0$ if $X_i=0$).

The integral above is $\lambda$-independent and, if it is convergent, the formula represents the correct asymptotic $T\gg U,\mu$ behavior of the integral, meaning that $W[C]\propto (T/U)^{\alpha_f/p}$. A careful treatment shows this is not always the case, as the integrand in \eqref{eq:tildeW_naive} can have ``flat directions'' \cite{SM}. These are flat directions which span the moduli space $\mathcal{M}_f(C)$ of the MFIS problem associated with $C$. The contributions that arise from letting $y_i\sim \lambda$ in these directions have been mistakenly neglected in the steps that led to \eqref{eq:tildeW_naive}---in particular, by letting the lower limits of integration $-\lambda X_i$ relax to $-\infty$ when $X_i>0$. We therefore expect that $\widetilde W[C]$ is bounded from above by an expression $\sim T^{\alpha_f/p} (\log T)^{\dim \mathcal M_f(C)}$. A similar treatment of the lower bound yields the same $T$-dependence and hence
\begin{equation}\label{eq:Wtilde_asympt}
    \widetilde W[C]\approx \xi(C) \left(\frac{T}{U}\right)^{\alpha_f(C)/p}\biggl(\log\frac{T}{U}\biggr)^{\dim \mathcal{M}_f(C)}\,,
\end{equation}
where $\xi(C)$ is a graph-dependent constant which scales at most exponentially in $N=|C|$, the number of vertices of the graph.

This is the correct asymptotic formula that we will use throughout the remainder of the paper.

\section*{Analysis of high-$T$ behavior}

Let us analyze the large-$T$ behaviour of a finite graph $G$. As we saw in \eqref{eq:partition_function_into_sectors}, the partition function decomposes into sectors identified by the vertex-induced subgraphs $\widetilde{G}$, whose weights $W[\widetilde{G}]$ can be evaluated in the large $T\to\infty$ limit by combining \eqref{isolated_vertex_asymptotic} and
\eqref{eq:Wtilde_asympt}. If we distinguish the subset of isolated vertices $\widetilde{G}_{\rm iso}$ from the nonisolated part $\widetilde{G}_{\text{non-iso}}$ of $\widetilde{G}$, we can write
$\widetilde G = \widetilde{G}_{\text{non-iso}} \sqcup \widetilde{G}_{\rm iso}$.
According to \eqref{isolated_vertex_asymptotic} and
\eqref{eq:Wtilde_asympt}, we have that the $W[\widetilde{G}]$ scales as (ignoring the subleading $\log T$ corrections)
\begin{equation}
    W[\widetilde G] \sim T^{\Gamma(\widetilde{G})},\,
    \label{eq:asympt_Gt_Giso}
\end{equation}
where the exponent
\begin{align}
 \Gamma(\widetilde G)&=\frac{\alpha_f(\widetilde{G}_{\text{non-iso}})}{p} + |\widetilde{G}_{\rm iso}|
 \label{exponent_1}
\end{align}
can also be rewritten, by using $\alpha_f(\widetilde G)=\alpha_f(\widetilde G_{\text{non-iso}})+|\widetilde G_{\rm iso}|$, as
\begin{align}
 \Gamma(\widetilde G)&=\frac{\alpha_f(\widetilde{G})}{p} + \left(1-\frac1p\right)|\widetilde{G}_{\rm iso}|.
 \label{exponent_2}
\end{align}

For a given value of $p$, we find the sector that dominates the partition function \eqref{eq:partition_function_into_sectors} at high temperature by maximizing $\Gamma(\widetilde G)$ over all vertex-induced subgraphs $\widetilde G\subseteq G$. 

From \eqref{exponent_1}, we see that there is a tension between maximising the two terms. 
On the one hand, $\alpha_f(\widetilde{G}_{\text{non-iso}})$, being linearly proportional to $|\widetilde{G}_{\text{non-iso}}|$, is maximised by taking $\widetilde{G}_{\text{non-iso}}$ as large as possible (i.e. $\widetilde G=G$), whereas $|\widetilde{G}_{\rm iso}|$ is maximised by taking $\widetilde{G}_{\text{non-iso}}$ as small as possible.
We can disentangle this tension by varying $p$. For $p\rightarrow 1$, we have that $\Gamma(\widetilde G)=\alpha_f(\widetilde G)$. Since $\alpha_f(\widetilde G)$ is maximized (perhaps non-uniquely) by taking $\widetilde G=G$, for $p\gtrsim1$ we expect that
\begin{equation}\label{eq:W_MFIS}
        W[G]\sim T^{\alpha_f(G) /p}, 
\end{equation}
and hence the dominant phase at high $T$ is the MFIS-gas.\footnote{The name MFIS-gas is not to be taken literally as we do not check that such a phase is always a gas. Indeed, for any bipartite lattice $\alpha=\alpha_f$ and so for $p=1$ the system can be, depending on the values of $U,\mu$, in either a MIS-solid or a MFIS-gas phase. We will comment more below.}

In the opposite limit of $p$ very large, the dominant configurations are those which maximize $|\widetilde{G}_{\rm iso}|$. As the set of isolated vertices $\widetilde{G}_{\rm iso}$ is an independent set of $\widetilde{G}\subseteq G$, its size $|\widetilde{G}_{\rm iso}|\le \alpha(G)$ is only maximised by choosing $\widetilde{G}$ to be one of the maximum independent sets $G_{\text{MIS}}$ of $G$. This results in
\begin{equation}
    W[G_{\rm MIS}]\sim T^{\alpha(G)}\;.
    \label{gMIS_phase}
\end{equation}

On the square lattice, this phase is precisely the checkerboard-solid discussed in \cite{Han:2025eiw,Huang:2025gvi}, so we call it the MIS-solid phase. Relations \eqref{gMIS_phase} and \eqref{eq:W_MFIS} lead immediately to the equipartition theorem of Fig.~\ref{fig:equipartition} (the case $p<1$ is reserved for \cite{SM}).

{
So far we have not discussed the stability of these high-$T$ phases for large but finite $T$. 
The stability is graph-dependent (e.g. the 1d lattice will not have a stable phase). 
However, note that the equipartition relations of Fig.~\ref{fig:equipartition} do not depend on the stability of the phases. 
We will review the Peierls argument on a square lattice for completeness to establish the stability in this case. In preparation for this, we first discuss bipartite graphs.}

\subsection{Bipartite graphs}

For any bipartite graph $G$, it is the case that $\alpha(G)=\alpha_f(G)$. This also holds for any vertex-induced subgraph $\widetilde G \subseteq G$.
Hence, we have the inequality $\alpha_f(\widetilde G) = \alpha(\widetilde G) \le \alpha(G)$. 
On the other hand, we also have that $|\widetilde G_{\rm iso}|\le \alpha(G)$, and so \eqref{exponent_2} becomes
\begin{equation}
    \Gamma(\widetilde G) = \frac{\alpha({\widetilde G})}{p} + \left(1-\frac1p\right)|\widetilde{G}_{\rm iso}| 
    \le \alpha(G)
    \label{inequalities_bipartite}
\end{equation}
This leads us to the conclusion that on the bipartite lattice, the leading contribution at high $T$ is given by $Z\sim T^{\alpha(G)}$. Further, when $p>1$ the dominant configurations (subject to Peierls stability) are given by any subgraph $\widetilde G$ for which $|\widetilde G_{\rm iso}|=\alpha(G)$. These configurations correspond to the MIS-solid phase, with the equipartition theorem $\avg{H}=\alpha(G)T$. 

The case of $p=1$ is more subtle, and the remainder of this section is dedicated to it. In this case, to saturate the inequality \eqref{inequalities_bipartite}, we must have $\alpha(\widetilde G)=\alpha(G)$. It is not difficult to see that for this to hold we must have that the MIS of $\widetilde G$ coincides with that of $G$, i.e. that $\widetilde G_{\rm MIS}=G_{\rm MIS}$. Therefore, at least one of the MIS of $G$ must be contained in $\widetilde G$, i.e. there must exist $G_{\rm MIS}$ such that $G_{\rm MIS}\in \widetilde G$. A general such configuration $\widetilde G$ is that of a $G_{\rm MIS}$, decorated by activating nodes in $G$ which are not in $G_{\rm MIS}$. We can think of this as the MIS-solid configuration decorated with the MFIS gas.

Let us now consider the ratio of weights of the total MIS-solid $\widetilde G=G_{\rm MIS}$ to that of the total MFIS-gas $\widetilde G=G$
\begin{equation}\label{eq:WGmisdivWG}
    \frac{W[G_{\rm MIS}]}{W[G]}\propto \left(\gamma\frac{U}{\mu}\right)^{\alpha(G)}\frac{1}{\log T}\;,
\end{equation}
where $\gamma$ is a dimensionless constant we are unable to compute with current methods.%
\footnote{The constant $\gamma$ in principle depends on $U/\mu$, but is finite in the limit $U\rightarrow \infty$, \cite{SM}.} The $\log T$ factor occurs when $G$ admits only two maximum independent sets (e.g. when $G$ is a regular square lattice with an even number of vertices in each direction).
 
Notice that this ratio goes to zero when $T\rightarrow\infty$ because of the $\log$, superficially implying that MFIS-gas always dominates if the $\log T$ term is present. However, the ratio depends on $\alpha(G)$ which scales as the system size $|G|$. If $U\gg \mu/\gamma$, the ratio above blows up in the thermodynamic limit, implying that in fact the MIS-solid dominates and is the correct thermodynamic phase for large $T\gg U\gg \mu$ (subject to Peierls' argument, which we review in the SM \cite{SM} for the square lattice), consistent with the findings of \cite{Huang:2025gvi}.

\begin{figure}
    \centering
    \def\svgwidth{\columnwidth}
    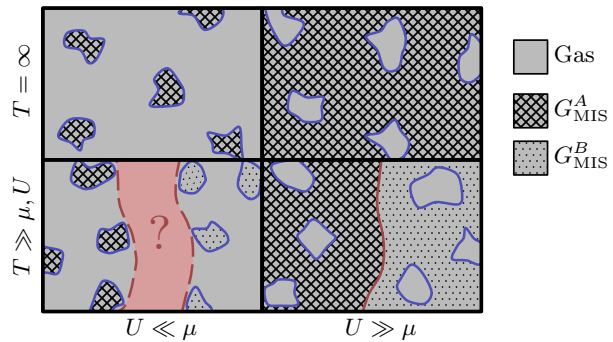
    \caption{A sketch of configurations for $p=1$ when $T\rightarrow\infty$ and $T$ large but finite.}
    \label{fig:p1_configs}
\end{figure}

For the remainder of this section, let $G$ be a 2d square $L\times L$ lattice. In that case, the MIS configurations are the two checkerboard states where $s_i=1$ on sublattice $A$ and zero on the sublattice $B$ and vice versa. We label these $G^A_{\rm MIS}$ and $G_{\rm MIS}^B$ respectively. As explained above, the configurations $\widetilde G$ which obey the same dominant high-$T$ scaling $W[\widetilde G]\sim T^{\alpha(G)}=T^{L^2/2}$ must have as subgraphs either $G_{\rm MIS}^A\subseteq \widetilde G$ or $G_{\rm MIS}^B\subseteq \widetilde G$.

When $U\gg \mu$, the dominant $T\rightarrow\infty$ configurations are those of $G_{\rm MIS}^A$ or $G_{\rm MIS}^B$, decorated by pockets of MFIS gas\footnote{One may worry that pockets of gas may have additional $\log(T)$ but it is not difficult to see that this never happen, see \cite{SM}.}. These pockets of gas are suppressed by a power of $(\mu/U)^{A}$ where $A$ is the size of the gas pocket (see Fig.~\ref{fig:p1_configs}). 

As $U$ decreases, these pockets of gas become bigger and eventually proliferate. However, note that in the limit $T\rightarrow\infty$, configurations which host both $G_{\rm MIS}^A$ and $G_{\rm MIS}^B$ are suppressed, and so long-range order persist for all $U>0$. But it is questionable whether the strict $T\rightarrow\infty$ limit has any meaning, as the partition function is not well defined in this limit. We therefore have to study the finite but large temperature regime.

When $U\gg\mu$ the standard Peierls argument follows, and the large-$T$ phase is a solid. This is reviewed in \cite{SM}. However, when $ U \ll \mu $ the domain walls are thick, and Peierls argument is difficult to make cleanly. Assuming a gapped phase, for any fixed $U$ we can always take the temperature to be large enough and cause the tension to be high, indicating that the system orders entropically, for any $U/\mu$, at sufficiently high $T$. Yet Monte Carlo and ``large color'' results of \cite{Huang:2025gvi} give evidence that for $U\lesssim \mu/2$ order disappears for $T\rightarrow\infty$. Nevertheless the discussion of \cite{Huang:2025gvi} may not extend all the way to $T\rightarrow \infty$. Further, the large color analysis indicates the high temperature phase may be gapless, invalidating our assumptions above. We leave the question of the high-$T$ phase in the $U\ll\mu$ regime for future discussions.

\subsection{General Graphs and Glassy Phases}

On a general graph, the dominant contribution for $p\ge 1$ has to maximize the exponent of $T$ in \eqref{eq:asympt_Gt_Giso}
\begin{equation}
    \Gamma_{\text{max}}(p)=\max_{\widetilde G\subseteq G}\Gamma(G)
\end{equation}
When $p\gg 1$, we must maximize $|\widetilde G_{\rm iso}|$, which is clearly maximized when $\widetilde G=G_{\text{MIS}}$.  

So \emph{for any graph $G$, the system solves a MIS optimization problem if temperature is high enough and $p$ is large enough.} Since MIS is an NP-complete problem on generic graphs, the strong Church-Turing thesis implies that the high temperature phase of such a system is glassy---it will require an exponentially long time in the system size to thermalize.

But now consider a large non-bipartite connected graph $G$, e.g. a triangular lattice, and define the model \eqref{hamiltonian} on it. This model is not glassy, as the MIS is easily found by populating $1/3$ of the triangular lattice. It is not difficult to see that this is the correct high-$T$ phase as long as $p>3/2$. However, consider a setup where we randomly set $U=0$ on some of the links. This reduces the model on $G$ to a model on its subgraph $H$ which is obtained by removal of the corresponding links. The MIS problem on such graphs will in general be NP-hard, and so we expect that the systems with some degree of ``link vacancies'' will experience glassy behavior. Moreover, we can make $U>0$ link-dependent and randomly distributed. Such models will in general be solving a complicated integer linear programming problem when $T$ is large. These problems are often NP-hard, resulting in the corresponding glassy phase. We leave detailed explorations of these \emph{entropic glass} phases for the future.

\section*{Acknowledgements}

We thank Stefano Cremonesi, Jeffrey Giansiracusa, Tyler Helmuth, Xiaoyang Huang, Zohar Komargodski, Costantinos Papageorgakis, Fedor Popov. This work is supported by the URF grant of the Royal Society of London and the
STFC Consolidated Grant ST/X000591/1.

\bibliography{refs.bib}

\makeatletter 
    
\renewcommand\onecolumngrid{
\do@columngrid{one}{\@ne}
\def\set@footnotewidth{\onecolumngrid}
\def\footnoterule{\kern-6pt\hrule width 1.5in\kern6pt}
}

\renewcommand\twocolumngrid{
        \def\footnoterule{
        \dimen@\skip\footins\divide\dimen@\thr@@
        \kern-\dimen@\hrule width.5in\kern\dimen@}
        \do@columngrid{mlt}{\tw@}
}

\makeatother    

\newpage
{
\setcounter{equation}{0}
\renewcommand{\theequation}{SM\arabic{equation}}
\onecolumngrid
\begin{appendix}
    \section{\Large Supplemental Material}

\section{Rigorous proof of the asymptotic formula}

This section is a rigorous treatment of formula \eqref{eq:Wtilde_asympt} for the large $T$ asymptotics of $\widetilde W[C]$, where $C$ is a connected graph consisting of $N\geq 2$ vertices. 
We will establish this formula by bounding $\widetilde W[C]$ from both above and below by the same asymptotic function of $T$ as $T\rightarrow\infty$. In this section we restrict our attention to $p\ge1$ for concreteness, but similar analysis can be applied to the case $p<1$ as we discuss briefly in the next section.

\subsubsection{Upper bound}

We begin with the proof of the upper bound ($K=\dim \mathcal M_f(C)$)
\begin{equation}\label{eq:upperbound}
    \widetilde W[C]\le \xi_+(C)\biggl(\frac{T}{U}\biggr)^{\alpha_f/p} \biggl(\log \frac{T}{U} \biggr)^K,
\end{equation}
where $\xi_+(C)$ is a graph-dependent constant.

The exponent of the integrand in \eqref{integral_y} involves a sum over all bonds, but if inactive bonds are dropped from the sum, then we obtain an upper bound on the integral.
A further upper bound is obtained by dropping the terms involving $\mu$.
Thus
\begin{equation}\label{eq:I_inequality}
\widetilde W[C]
\le \frac{e^{ \lambda \alpha_f/p} }{p^N} \biggl[ \prod_i\int_{-\lambda X_i}^\infty d y_i \biggr] \; f(\vec y)\,\,,
\end{equation}
where
\begin{equation}
\label{eq:integrand_f} 
f(\vec y)=\exp \biggl(\vec a \cdot \vec y/p -\sum_{\avg{ij} \; \mathrm{active}} e^{\vec b_{ij} \cdot \vec y} \biggr),
\end{equation}
and $\vec a = \vec e_1 + \ldots + \vec e_N$, $\vec b_{ij} = \vec e_i + \vec e_j$. Here, $(\vec e_j)_k=\delta_{jk}$ denote the standard orthonormal basis vectors of $\mathbb R^N$.

To proceed, we decompose $\vec y$ into components tangent and normal to $\mathcal{M}_f(C)$, $\vec y = \vec y_\parallel + \vec y_\perp$; 
the tangent and normal spaces will be denoted $V^\parallel$ and $V^\perp$.%
\footnote{
As is standard practice, we treat tangent and normal spaces as linear subspaces of the ambient Euclidean space.
}
Note that $\vec a \cdot \vec y_\parallel = 0$ and $\vec b_{ij} \cdot \vec y_\parallel = 0$ for all active bonds.
In particular, this implies that $f(\vec y)=f(\vec y_\perp)$. 

For fixed $\vec y_\perp$, the integral over $\vec y_\parallel$ is simply the $K$-dimensional volume of the part of the $K$-dimensional affine subspace $\vec y_\perp + V^\parallel$ inside the total integration domain $\prod_i [-\lambda X_i,\infty)$.
Clearly, this volume scales at most as $\lambda^K$ with a coefficient $g(\vec y_\perp)$ depending polynomially on $\vec y_\perp$. 
The situation is sketched in Fig.~\ref{fig:integration_limits}.
Thus
\begin{equation}
    \widetilde W[C] \leq \frac{e^{\lambda \alpha_f/p} \lambda^K}{p^N} \int_{\mathcal D} d^{N-K} y_\perp \; g(\vec y_\perp)f(\vec y_\perp),
\end{equation}
where $\mathcal D$ denotes the integration domain of $\vec y_\perp$. 
\begin{figure}
    \centering
    \def\svgwidth{.8\columnwidth}
    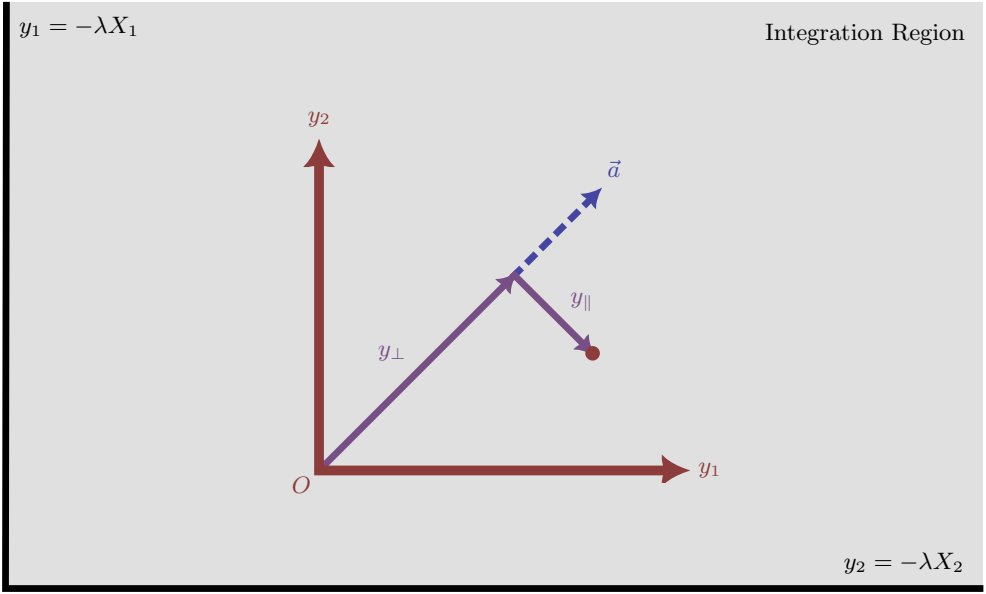
    \caption{
    At fixed $\vec y_\perp$, the integration over $\vec y_\parallel$ sweeps over a $K$-dimensional subspace of volume at most $\lambda^K$ times a polynomial in $\vec y_\perp$. 
    (In the figure $K=1$.)
    }
    \label{fig:integration_limits}
\end{figure}
This is the desired upper bound on $\widetilde W[C]$ provided the last integral over $\vec y_\perp$ converges even as $\lambda \to \infty$.

Note that along any direction $\vec y_\perp$ with $\vec a \cdot \vec y_\perp < 0$, $f(\vec y_\perp)$ decays exponentially, and along any direction $\vec y_\perp$ with $\vec b_{ij} \cdot \vec y_\perp > 0$ for at least one active bond $\langle ij \rangle$, $f(\vec y_\perp)$ decays super exponentially.
The ``bad directions'' along which $f(\vec y_\perp)$ does not decay at least exponentially fast are characterized by the condition:
\begin{equation}
    \vec a \cdot \vec y_\perp \geq 0 \ \text{and} \ \vec b_{ij} \cdot \vec y_\perp \leq 0 \ \text{for any active bond} \ \langle ij \rangle.
    \label{bad-direction}
\end{equation}

Let us prove that any such bad direction $\vec y_\perp$ is not in the integration domain $\mathcal D$.
This means we must show that for any $\vec y_\parallel$, we have $\vec y = \vec y_\parallel + \vec y_\perp \notin \prod_i [-\lambda X_i,\infty)$.
To this end, consider the vector $\vec x(\epsilon) = \vec X + \epsilon \vec y$.
Since $\vec y_\perp$ satisfies \eqref{bad-direction} and $\vec a \cdot \vec y_\parallel=0$, we have $\vec a \cdot \vec x(\epsilon) \geq \vec a \cdot \vec X = \alpha_f$.
Now, $\vec y$ is not tangent to $\mathcal M_f(C)$, so $\vec x(\epsilon)$ is either inside $\mathcal A(C)-\mathcal M_f(C)$ or outside of $\mathcal A(C)$. 
But since $\alpha_f$ is the maximum of the quantity $\vec a \cdot \vec x$ over $\vec x \in \mathcal A(C)$, we must have $\vec x(\epsilon) \notin \mathcal A(C)$.

The fact that $\vec x(\epsilon) \notin \mathcal A(C)$ means $\vec x(\epsilon)$ must violate at least one of the inequalities defining $\mathcal A(C)$; in other words, at least one of the following three conditions must hold:
\begin{itemize}
    \item[(i)] For some $i$, $x_i(\epsilon)<0$.
    \item[(ii)] For some $i$, $x_i(\epsilon)>1$.
    \item[(iii)] For some $\langle ij \rangle$, $x_i(\epsilon) + x_j(\epsilon)>1$. 
\end{itemize}
If (i) holds, then clearly $X_i=0$ and $y_i<0$, and it follows that $y_i \notin [0,\infty) = [-\lambda X_i,\infty)$, which is what we wanted to show.
If (ii) holds, then clearly $X_i=1$ and $y_i>0$.
But for any given neighbor $j$ of $i$, we must have $X_j=0$ as well as that $\langle ij \rangle$ be an active bond.
It then follows from \eqref{bad-direction} that $y_i + y_j = \vec b_{ij} \cdot \vec y_\perp \leq 0$, whence that $y_j \leq - y_i < 0$, and hence that $y_j \notin [0,\infty) = [-\lambda X_j,\infty)$, again as we wanted to show.
Finally, if (iii) holds, then clearly $X_i + X_j = 1$ and $y_i + y_j > 0$.
This means that $\langle ij \rangle$ is an active bond, so by \eqref{bad-direction}, $y_i + y_j = \vec b_{ij} \cdot \vec y_\perp \leq 0$, which is a contradiction.
Thus, (iii) can never hold. 
We have shown that in all three cases, $\vec y \notin \prod_i [-\lambda X_i,\infty)$, and this completes the proof that $\vec y_\perp \notin D$.

We will now establish the fact that there is a constant $c>0$ such that for any $\vec y_\perp \in \mathcal D$, at least one of the conditions is true:
\begin{itemize}
    \item[(a)] $\vec a \cdot \vec y_\perp \leq - c |\vec y_\perp|$,
    \item[(b)] $\vec b_{ij} \cdot \vec y_\perp \geq c |\vec y_\perp |$ for some active bond $\langle ij \rangle$. 
\end{itemize}
Indeed, this fact allows us to bound the magnitude of $f(\vec y_\perp)$ everywhere in $\mathcal D$ by a manifestly integrable function.
For if $\vec y_\perp$ satisfies (a), then
\begin{equation}
    f(\vec y_\perp) \leq \exp(- c|\vec y_\perp|/p)
\end{equation}
while if $\vec y_\perp$ satisfies (b), then
\begin{equation}
    f(\vec y_\perp) \leq \exp(|\vec a| \, |\vec y_\perp|/p - e^{c |\vec y_\perp|} ) 
\end{equation}
and the integrability of $f(\vec y_\perp)$ over $\mathcal D$ is manifest. 

We turn now to the proof that there is indeed a constant $c>0$ such that for any $\vec y_\perp \in \mathcal D$, either conditions (a) or (b) hold.
We will prove this by contradiction.
Suppose that for each $c>0$, there exists $\vec y_\perp \in \mathcal D$ such that (a) and (b) are false, i.e., such that
\begin{equation}
    \vec a \cdot \vec y_\perp > - c |\vec y_\perp| \ \text{and} \ \vec b_{ij} \cdot \vec y_\perp < c | \vec y_\perp | \ \text{for any active bond} \ \langle ij \rangle
\end{equation}
This means we can find a sequence of unit vectors $\hat u_n \in \mathcal D \cap S^{N-1}$ such that
\begin{equation}
    \vec a \cdot \hat u_n > - 1/n \ \text{and} \ \vec b_{ij} \cdot \hat u_n < 1/n \ \text{for any active bond} \ \langle ij \rangle
\end{equation}
Since $\mathcal D$ is closed, $\mathcal D \cap S^{N-1}$ is compact, so we may assume (after possibly passing to a subsequence) that the sequence $\hat u_n$ converges to a limit $\hat u_* \in \mathcal D \cap S^{N-1}$, which necessarily satisfies 
\begin{equation}
    \vec a \cdot \hat u_* \geq 0 \ \text{and} \ \vec b_{ij} \cdot \hat u_* \leq 0 \ \text{for any active bond} \ \langle ij \rangle
\end{equation}
In other words, $\hat u_*$ satisfies the condition of a bad direction, \eqref{bad-direction}.
But this contradicts the fact that the bad directions are outside of the integration domain $\mathcal D$.
This completes the proof, and hence the upper bound \eqref{eq:upperbound} is established.

\subsubsection{Lower bound}
By restricting the integration to the finite polytope $\mathcal A(C)$ given by \eqref{eq:Pdomain}, we obtain
\begin{equation}
    \widetilde W[C]
    \ge \frac{\lambda^N}{p^N}\int_{\mathcal A(C)} d^N x \; \exp\biggl({\frac{\lambda}{p}\vec a \cdot \vec x -\sum_{\avg{ij}}e^{\lambda(\vec b_{ij} \cdot \vec x -1)}-\frac{\mu}{U}\sum_i e^{(x_i/p-1)\lambda}}\biggr)
\end{equation}
Since for $\vec x \in \mathcal A(C)$, the quantities $\vec b_{ij} \cdot \vec x -1$ and $x_i/p-1$ are bounded from above, replacing them by their upper bounds (zero) further gives
\begin{equation}
\widetilde W[C] \ge \frac{\lambda^N e^{-N' - N\mu/U}}{p^N}\int_{\mathcal A(C)} d^N x\; e^{\lambda \vec a \cdot \vec x / p},
\label{lower_1_bound}
\end{equation}
where $N'$ is the number of bonds of $C$.
Let $\Omega$ be the set of all normal directions $\hat u_\perp \in V^\perp \cap S^{N-1}$ such that $\vec X + \epsilon \hat u_\perp \in \mathcal A(C)$ for some $\epsilon>0$.
Let us write $\vec x = \vec X + \vec z_\parallel + \vec z_\perp$ with $\vec z_\parallel \in V^\parallel,\ \vec z_\perp \in V^\perp$, and change integration variables to $\vec z_\parallel$ and $\vec z_\perp$.
It is easy to see that we can choose $\epsilon_\parallel,\epsilon_\perp>0$ small enough so that $\vec X + \vec z_\parallel + \vec z_\perp \in \mathcal A(C)$ provided $\vec z_\parallel,\vec z_\perp$ both satisfy $|\vec z_\parallel|<\epsilon_\parallel,\ |\vec z_\perp|<\epsilon_\perp/\lambda,\ \hat z_\perp \in \Omega$.
Thus, by further restricting the integration to such $\vec z_\parallel$ and $\vec z_\perp$, we obtain
\begin{equation}
    \widetilde W[C] \ge \lambda^N e^{\lambda \alpha_f/p} p^{-N} e^{-N' - N\mu/U} \int_{|\vec z_\parallel|<\epsilon_\parallel} d^K z_\parallel \int_{|\vec z_\perp|<\epsilon_\perp/\lambda,\, \hat z_\perp \in \Omega} d^{N-K} z_\perp \;  e^{\lambda \vec a \cdot \vec z_\perp / p} .
\end{equation}
Rescaling $\vec z_\perp \to \vec y_\perp/\lambda$, this becomes
\begin{equation}
    \widetilde W[C] \ge \lambda^K e^{\lambda \alpha_f/p} p^{-N} e^{-N' - N\mu/U} \int_{|\vec z_\parallel|<\epsilon_\parallel} d^K z_\parallel \int_{|\vec y_\perp|<\epsilon_\perp,\, \hat y_\perp \in \Omega} d^{N-K} y_\perp \;  e^{\vec a \cdot \vec y_\perp / p}.
    \label{eq:lowerbound}
\end{equation}
As the remaining integrals do not depend on $\lambda$, we can assemble them along with all the other $\lambda$-independent factors into a finite constant $ \xi_-(C)$.
Therefore,
\begin{equation}
    \widetilde W[C]\ge \xi_-(C) \left(\frac{T}{U}\right)^{\alpha_f/p} \biggl(\log \frac{T}{U} \biggr)^K \,.
    \label{below}
\end{equation}

Together with \eqref{eq:upperbound}, this implies
\begin{equation}
    \widetilde W[C] \sim \xi(C)\biggl(\frac{T}{U}\biggr)^{\alpha_f(C)/p} \biggl( \log \frac{T}{U} \biggr)^{\dim \mathcal M_f(C)} .
    \label{final}
\end{equation}
for a graph-dependent and temperature-independent coefficient $\xi(C)$.

Finally, we stress that to arrive at \eqref{below} or \eqref{eq:upperbound}, we have never used $p\ge1$. In fact, \eqref{final} holds also for $0<p<1$, upon replacing $\mathcal A(C),\alpha_f(C),\mathcal M_f(C)$ with the obvious corresponding quantities/objects, see the discussion following under formula \eqref{new_polytope}. 

\subsection{Scaling of $\xi(C)$}

Let us comment on the behavior of the coefficients $\xi(C)$. While this is not temperature dependent, the coefficient scales exponentially with the size of the graph $N=|C|$ as $N\rightarrow\infty$. Note that $\xi(C)$ has some dependence on $\mu/U$ that we will not worry about in the case of $p>1$, while in the case $p=1$ we will mostly be interested in the regime $U\gg \mu$ where the dependence drops out.

One way to argue this is to note that $\widetilde W(C)$ can be viewed as a partition function of local physical model on a graph $C$. The free energy of such models is expected to be extensive in the size $N$ of the system, and so $\widetilde W$ is at most exponential in $N$. A more quantitative statement is that $\widetilde W < \widetilde W|_{U=0}\propto \left(T/\mu\right)^N$, and so cannot scale faster than exponential. The issue with this argument is that the RHS of this inequality scales faster as $T\rightarrow\infty$ than that of $\widetilde W$, so one may worry about the order of limits $T\rightarrow \infty$ and $N\rightarrow\infty$. 

We can do better by noticing that $\xi(C)$ comes from the integration over $z_\beta$--directions on which the integral converges even when $\lambda\rightarrow\infty$, of which there are at most $N$. Let us briefly sketch the argument for the case $K=0$. In this case, the MFIS has no degeneracies and so we can simply consider the integral in \eqref{eq:tildeW_naive}. This integral extends over an infinite region, with some of the coordinates $y_i$ possibly bounded from below by $y_i=0$. We call this infinite region $R$. The integrand, which we label $h(y)$, has no singularities in the region of integration, and therefore attains its maximal value at some $y_i=Y_i$, i.e. $h_\text{max} \equiv h(Y)$. Further, because of the local nature of the exponent, the saddle point equation and the solution on $Y_i$ is not expected to be dramatically affected as we take the limit $N\rightarrow\infty$.

Moreover, $h(y)$ decays rapidly (at least exponentially) in all infinite directions of $R$. We can therefore bound the integral $\int \diff^N y\; h(y)$ from above by replacing the integrand with its maximal value inside a box $B$ contained in region $R$, i.e. $\int_R d^N y\; h(y)< \int_B dy^N h_\text{max}=V(B) h_\text{max}$, where $V(B)$ is the volume of the box\footnote{Since $h(y)\le h_\text{max}$, the integral in the region $\int_R h(r)\le \int_Bd^Ny h_\text{max}+\int_{R-B}d^Ny f(y)$. By taking $B$ to be sufficiently large, $\int_{R-B}f(y)d^Ny$ can be made arbitrarily small.} $B$. It takes little thought to see that $h_\text{max}$ is at most exponential in $N$ (provided that the coordination number of the graph does not scale with $N$), and since the same is true for $V(B)$, we have demonstrated that $\int_R d^N\;y h(y)\le \gamma^N$ for some $\gamma$.

Since $\alpha_f(C)$ scales linearly with $N$, $\xi(C)$ can be absorbed into the ratio $T/U$ such that $\widetilde W\sim (\gamma T/U)^{\frac{\alpha_f(C)}{p}}$ as $N\rightarrow \infty$. 
Note that while $\gamma$ depends on the ratio $\mu/U$, by \eqref{eq:upperbound} it is bounded above by a finite constant independent of $\mu/U$.
Moreover, by \eqref{eq:lowerbound}, $
\gamma$ will be become independent of $
\mu/U$ for $U\gg \mu$.

\section{Asymptotics for $p<1$}

Let $C$ be a connected graph with $N=|C|$ vertices and consider the integral $\widetilde W[C]$ as defined in \eqref{eq:WtildeP}. For any $p>0$, such integral is super-exponentially suppressed outside of the region of integration $\mathcal A_p(C)$ defined by
\begin{equation}
\label{new_polytope}
    \mathcal A_p(C) = \{ \vec x \in [0,p]^N \mid x_i+x_j\le 1 \;\; \forall \avg{ij}\in C\}.
\end{equation}
Hence, by neglecting $\log T$ terms, in the high-$T$ regime we have
\begin{align}
    \widetilde W[C]\approx T^{\frac{\alpha_p(C)}p}
\end{align} where $\alpha_p(C)=\max_{\mathcal A_p(C)}\sum_{i=1}^N x_i$, and $\alpha_p(C)/p$ is the equipartition coefficient depicted in Fig.~\ref{fig:equipartition} of the main text.

It is easy to see that for $p\ge1$, $\mathcal A_p(C)$ agrees with the domain $\mathcal A(C)$ defined in \eqref{eq:Pdomain}, in which case $\alpha_p(C)=\alpha_f(C)$ and $\widetilde W[C]\sim T^{\alpha_f(C)/p}$, as we discussed under \eqref{eq:Pdomain}.

It is also easy to see that for $p\le\frac12$ the edge constraints are redundant ($x_i+x_j\le 2p\le1$) so in this case $\mathcal A_p(C)=[0, p]^{\times N}$ is a hypercube of dimension $N$ and side $p$, for which $\alpha_p(C)=Np$; so in the high-$T$ limit all interactions are washed away for $p\le\frac12$, and the equipartition coefficient $\alpha_p(C)$ simply captures the number of decoupled sites.

For $\frac12\le p \le 1$, the domain $\mathcal A_p(C)$ interpolates between the polytope $\mathcal A(C)$ and the hypercube $[0, p]^{\times N}$, so $\alpha_p(C)$ will be a linear combination of the corresponding equipartition coefficients $\alpha_p(C)=\alpha_f(C)$ and $\alpha_p(C)=Np$. To see this, write $x_i=p-d_i$:
\begin{align}
    \max_{\mathcal A_p(C)}\sum_{i=1}^N x_i=Np-
\min_{\substack{
0 \le d_i \le p \\
d_i + d_j \ge 2p - 1
}}
\sum_{i=1}^N d_i
\label{expon_something}
\end{align}
Appreciate that when minimizing $\sum_i d_i$ for $d_i\ge0$ and $d_i + d_j \ge 2p - 1$, we automatically have that $d_i\le 2p-1$ for any $d_i$. Indeed, if there exists a $\hat{i}$ such that $d_{\hat{i}}>2p - 1$, then we could minimize $\sum_i d_i$ by substituting it with $d_{\hat{i}}\to d_{\hat{i}}'=2p - 1$ while keeping all the other $d_i$ unchanged, which is a feasible operation to do as $d_{\hat{i}}'+d_j=(2p-1)+d_j\ge (2p-1)$ for $d_j\ge0$. But for $p\in[\frac12,1]$, $d_i\le 2p-1\le p$ so \eqref{expon_something} reads as

\begin{align}
    \alpha_p(C)
&=Np-
\min_{\substack{
0 \le d_i  \\
d_i + d_j \ge 2p - 1
}}
\sum_{i=1}^N d_i\\
&=Np-(2p-1)
\min_{\substack{
0 \le d_i  \\
d_i + d_j \ge 1
}}
\sum_{i=1}^N d_i\\
&=Np-(2p-1)
\tau_f(C)
\\
&=N(1-p)+(2p-1)\alpha_f(C)
\end{align}
where we recognized the \emph{fractional minimum vertex cover number} $\tau_f(C)$, and 
used the standard identity $\tau_f(C)+\alpha_f(C)=N$, \cite{Karp1972}.

    \section{From the sum to the integral}

    Let us define the real and positive function
    \begin{equation}
        f(\{n_i\})=e^{-U\beta\sum_{\avg{ij}\in C}n_i^pn_j^p-\mu\beta\sum_{i\in C} n_i}
    \end{equation}
    and its sum
    \begin{equation}
    S(C)=\sum_{n_1=1}^\infty\sum_{n_2=1}^\infty\cdots \sum_{n_N=1}^\infty    f(\{n_i\})
    \end{equation}
    over a given connected graph $C$ such that $|C|\ge 2$. Analogously, for any $a>0$, we define the integral
    \begin{equation}
    I_a(C;\mu,U)=\int_a^\infty dn_1\int_a^\infty dn_2\cdots \int_a^\infty dn_N   f(\{n_i\})
    \end{equation}
    where we made dependence on $\mu$ and $U$ explicit. We will occasionally drop the arguments $\mu$ and $U$, however we will need this in what follows. 

Note that $I_{a}(C;\mu,U)=I_{1}(C;\mu a, U a^{2p})$. Along with the result \eqref{final}, we have that
\begin{equation}
    I_a(C)\approx a^{2\alpha_f(C)}\xi\left(C,\frac{U}{\mu}a^{2p-1}\right) \left(\frac{T}{U}\right)^{\frac{\alpha_f}{p}}(\log T/U)^{\dim \mathcal M_f(C)}
\end{equation}
which, shows that the high-$T$ behavior is the same for all $a$.

Now we will show that
\begin{equation}
    I_1(C,\mu,U)<S(C)<k^N I_{1-1/k}(C,\mu,U)
\end{equation}
for any $k>1$ and hence $S(C)$ must have the same high-$T$ asymptotics as $I_1(C)$.

To give the lower bound on $S(C)$, note that $f(\{n_i\})$ is a strictly decreasing function of all $n_i>0$, and therefore
    \begin{equation}
        f(\{n_i+\epsilon_i\})<f(\{n_i\})
    \end{equation}
    for $0<\epsilon_i<1$ and $n_i\ge0$. By integrating both sides of this inequality over $\epsilon_i$ on the unit box $\epsilon_i\in(0,1)$, and summing over all natural numbers $n_i\ge1$ we get the lower bound
    \begin{equation}
        I_1(C) < S(C)\;.
        \label{lower_bound_integral_sum}
    \end{equation}

    Now we show the more tricky, lower bound, noting that
    \begin{equation}
        f(\{n_i-\epsilon_i\})>f(\{n_i\})
    \end{equation}
    for $0<\epsilon_i< 1$ and $n_i\ge1$. We fix $k>1$. Integrating $\epsilon_i$ over the box $B_k: \epsilon_i\in(0,1/k]$ and summing over all natural numbers $n_i\ge1$ we obtain
    \begin{equation}
        \sum_{\{n_i=1,\cdots\}}\int_{B_k}d^N\epsilon f(\{n_i-\epsilon_i\})> \frac{1}{k^N}S(C)\;.
    \end{equation}
  The LHS is the integral over $n_i$ in region $R_k=\bigcup_{\{m_i\}}^\infty [m_1-1/k,m_1]\times [m_2-1/k,m_2]\times \cdots\times [m_N-1/k,m_N]$, i.e. 
    \begin{equation}
        k^N\prod_{i=1}^N\int_{R_k} d^N n\; f(\{n_i\})> S(C)\;.
    \end{equation}
    Notice that the integral on the LHS is always smaller than $I_{1/k}(C)$ because the integration region of $I_{1/k}(C)$ includes the region $R_k$, so
    \begin{equation}
        k^N I_{1-1/k}(C)> S(C)\;.
        \label{upper_bound_integral_sum}
    \end{equation}

    This completes our proof.

\section{The Peierls argument on the square lattice}

\begin{figure}[b]
    \centering
    \def\svgwidth{.85\columnwidth}
\begingroup%
  \makeatletter%
  \providecommand\color[2][]{%
    \errmessage{(Inkscape) Color is used for the text in Inkscape, but the package 'color.sty' is not loaded}%
    \renewcommand\color[2][]{}%
  }%
  \providecommand\transparent[1]{%
    \errmessage{(Inkscape) Transparency is used (non-zero) for the text in Inkscape, but the package 'transparent.sty' is not loaded}%
    \renewcommand\transparent[1]{}%
  }%
  \providecommand\rotatebox[2]{#2}%
  \newcommand*\fsize{\dimexpr\f@size pt\relax}%
  \newcommand*\lineheight[1]{\fontsize{\fsize}{#1\fsize}\selectfont}%
  \ifx\svgwidth\undefined%
    \setlength{\unitlength}{328.67198794bp}%
    \ifx\svgscale\undefined%
      \relax%
    \else%
      \setlength{\unitlength}{\unitlength * \real{\svgscale}}%
    \fi%
  \else%
    \setlength{\unitlength}{\svgwidth}%
  \fi%
  \global\let\svgwidth\undefined%
  \global\let\svgscale\undefined%
  \makeatother%
  \begin{picture}(1,0.43122731)%
    \lineheight{1}%
    \setlength\tabcolsep{0pt}%
    \put(0,0){\includegraphics[width=\unitlength,page=1]{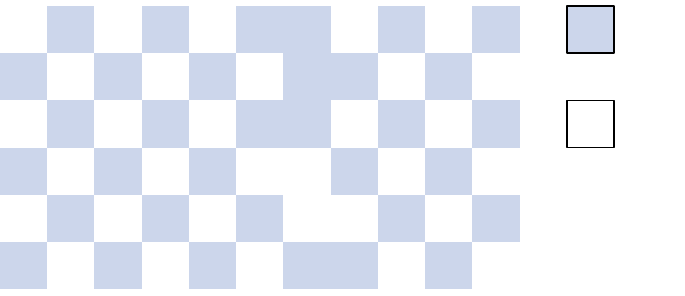}}%
    \put(0.90460092,0.23830458){\color[rgb]{0,0,0}\makebox(0,0)[lt]{\lineheight{1.25}\smash{\begin{tabular}[t]{l}$s_i=0$\end{tabular}}}}%
    \put(0.90460092,0.37629726){\color[rgb]{0,0,0}\makebox(0,0)[lt]{\lineheight{1.25}\smash{\begin{tabular}[t]{l}$s_i=1$\end{tabular}}}}%
    \put(0,0){\includegraphics[width=\unitlength,page=2]{dw_square.pdf}}%
  \end{picture}%
\endgroup%

    \caption{Domain wall between two MIS-solid checkerboard phases.}
    \label{fig:dw_square}
\end{figure}

In this section we specialize to the square lattice and demonstrate that the solid phase is stable when $p\ge 1$ by making a standard Peierls argument. 

The MIS-solid phase configuration is a checkerboard (on the dual lattice) and its weight scales as
\begin{equation}
    W[G_{\rm MIS}]\sim \left(\frac{T}{\mu}\right)^{\frac{L^2}{2}}\;.
\end{equation}

Now consider a domain wall configuration like the one in Fig.~\ref{fig:dw_square}. This induces a graph we will call $G_{\rm DW}$. 
The ratio of the domain wall configuration and the $W[G_{\rm MIS}]$ is given by
\begin{equation}
    \frac{W[G_{\rm DW}]}{W[G_{\rm MIS}]}\propto \left(T\right)^{-\gamma}\;.
\end{equation}
Since this ratio picks ups a contribution only along the domain wall, we must have that $\gamma\ge c\ell$ where $c>0$ is a fixed constant and $\ell$ is the length of the domain wall\footnote{Similar arguments can be made in higher dimensions.}. 

The dominant contribution to the partition function can therefore be expanded around a given MIS state as 
\begin{equation}
    Z\approx W[G_{\rm MIS}]+\sum_{\rm DW} W[G_{\rm DW}]=W[G_{\rm MIS}]\left(1+\sum_{G_{\rm DW}}\# e^{-\gamma[G_{\rm DW}] \log T}\right),
\end{equation}
where the sum is over all domain-wall configurations $G_{\rm DW}$ and where we indicated explicitly that the coefficient $\gamma$ is dependent on the given domain-wall. Since any domain-wall configuration differs from the checkerboard bulk by graphs supported along the length of the domain wall, we must have that $\gamma[G_{\rm DW}]$ is bounded from below by $\gamma_{G_{\rm DW}}\ge \Gamma\ell$ where $\ell$ is the domain-wall length and $\Gamma$ is domain wall independent.

The only way that domain walls can change a checkerboard phase is if the number of domain walls of fixed, but large, length $\ell$ scales faster than their weight. It is well known that the number of domain-walls of fixed length $\ell$ scales as\footnote{One way to approximate this on a square lattice is by a closed, non-backtracking random walk. A rough approximation gives that at each step there are $3$ directions, so an $\ell$-step 2d random walk has degeneracy of $3^\ell$. The return probability is roughly $1/\ell$, and so the closed non-backtracking random walk has a rough degeneracy of $3^\ell/\ell$.} $e^{\Gamma_0\ell}$ when $\ell$ is large ($\Gamma_0$ is temperature independent). Hence, a contribution of all domain walls of fixed length $\ell$ is bounded from above by $e^{\ell (\Gamma_0-g\log T)}$, which makes domain walls harmless as long as $T\gg e^{\frac{\Gamma_0}{g}}$ and the phase is stable. This shows stability for high, but not infinite, $T$.

The argument so far is for $p>1$. For $p=1$ the argument is the same as long as $U\gg \mu$. But it is not clear to us if the Peierls' argument goes through for $\mu\gtrsim U$, leaving open the possibility of the high-$T$ ordered phase for all values of $U>0$.

\section{$p=1$ and pockets of gas on a square lattice}

We have seen in the main text that on the square lattice $G$ that the most dominant configurations at large $T$ are those associated with a vertex-induced subgraph $\tilde G\subseteq G$ for which $\alpha_f(\tilde G)=\alpha_f(G)$. Here we show that all such configurations have a unique MFIS solution, except when $\tilde G=G$. This means that all dominant configurations have no $\log T$ factors in their weight.

As stated in the main text, on a bipartite square lattice $G$, the dimension of the MFIS moduli space is $1$.
The MFIS solutions $\{X_i\}$ are given by $X_i = t$ for $i \in G^A_{\text{MIS}}$ and $X_i=1-t$ for $i \in G^B_{\text{MIS}}$ for any $t \in [0,1]$.
Here, we prove that on any other vertex-induced subgraph $\widetilde G$ of $G$ with $\alpha_f(\widetilde G) = \alpha_f(G)$, there is precisely one MFIS solution. 

We will make use of the fact that for any MFIS solution $\{x_i\}$ on $\widetilde G$, there corresponds a solution $\{X_i\}$ to the MFIS problem on $G$ given by assigning $X_i = x_i$ to $i \in \widetilde G$ and $X_i = 0$ to $i \notin \widetilde G$.
Indeed, our assumption that $\alpha_f(\widetilde G) = \alpha_f(G)$ trivially implies
\begin{equation}
    \sum_{i \in G} X_i = \sum_{i \in \widetilde G} x_i = \alpha_f(\widetilde G) = \alpha_f(G)
\end{equation}
so $\{X_i\}$ is indeed a valid MFIS solution on $G$. 

We have stated in the main text that on a bipartite graph, the only way a vertex-induced subgraph $\widetilde G \neq G$ can have $\alpha_f(\widetilde G) = \alpha_f(G)$ is for $\widetilde G$ to contain either $G^A_{\text{MIS}}$ or $G^B_{\text{MIS}}$ (but not both, of course).
So without loss of generality, suppose $\widetilde G$ contains $G^A_{\text{MIS}}$ but not $G^B_{\text{MIS}}$.

Clearly, one MFIS solution on $\widetilde G$ is to assign $x_i=1$ to $i \in G^A_{\text{MIS}}$ and $x_i=0$ to $i \in \widetilde G \cap G^B_{\text{MIS}}$. 
To prove there are no other MFIS solutions on $\widetilde G$, suppose for sake of contradiction that $\{x_i\}$ is some other MFIS solution.
Then on some vertex $i \in \widetilde G \cap G^B_{\text{MIS}}$, we must have $x_i>0$.
Thus, the associated MFIS solution $\{X_i\}$ on $G$ satisfies $X_i>0$ on this vertex $i \in G^B_{\text{MIS}}$.
Further, by hypothesis, there is a vertex $j \in G^B_{\text{MIS}}$ that is not in $\widetilde G$, so for this vertex $j$, we have $X_j = 0$.
But this contradicts the fact any MFIS solution on $G$ must have all $X$s on $G^B_{\text{MIS}}$ equal.
Thus, we have shown that there is precisely one MFIS solution on $\widetilde G$. 

In particular this means that for the $p=1$ model on the square lattice, the Boltzmann weight of any MIS-solid state decorated by pockets of MFIS-gas has no $\log T$ enhancement over the pure MIS-solid state.

\end{appendix}

}

\end{document}